\documentclass[10pt,a4paper]{article}
\RequirePackage{latexsym,amsmath,amssymb}
\RequirePackage[dvipsnames,usenames]{color}

\usepackage{fullpage}

\usepackage[english]{babel}
\usepackage[T1]{fontenc}
\usepackage[final]{showkeys} 
\usepackage[hidelinks]{hyperref}
\hypersetup{pdfauthor={M. Cadoni, M. Tuveri...},
            pdftitle={}
            }
\usepackage[]{cleveref}
\usepackage{amsmath}
\usepackage{ mathrsfs }
\usepackage{subcaption}
\usepackage{cancel}
\usepackage{url}

\usepackage{tikz}
\usetikzlibrary{decorations.pathmorphing}

\DeclareMathOperator{\Tr}{Tr}
\Crefname{equation}{Eq.}{Eqs.}
\Crefname{figure}{Fig.}{Figs.}
\crefformat{plural}{#2eqs.~(#1)#3}
\crefname{section}{Sect.}{Sects.}

\usepackage{amsfonts}
\usepackage{graphicx,epsfig}
\def\be#1\ee{\begin{align}#1\end{align}}

\renewcommand{\ge}{\geqslant}
\renewcommand{\geq}{\geqslant}
\renewcommand{\le}{\leqslant}

\def\lb{\label}

\def\beq{\begin{equation}}
\def\eeq{\end{equation}}
\title{\bf Unitarity and  Page curve for  evaporation of  2D  AdS black holes}
\author{M.~Cadoni${}^{ab}$\thanks{E-mail: mariano.cadoni@ca.infn.it},
A. P. ~Sanna${}^{ab}$\thanks{E-mail: asanna@dsf.unica.it} \ 
\\
${}^a$\emph{Dipartimento di Fisica, Universit\`a di Cagliari}
\\
{\em Cittadella Universitaria, 09042 Monserrato, Italy}
\\
\\
${}^b$\emph{I.N.F.N, Sezione di Cagliari}
\\
{\em  Cittadella Universitaria, 09042 Monserrato, Italy}
\\
\\}
\begin{document}
\maketitle
\begin{abstract}
We explore the Hawking evaporation of  two-dimensional anti-de  Sitter  (AdS$_2$), dilatonic black hole  coupled  with conformal matter, and derive the Page curve for the entanglement entropy  of radiation.
We first  work in a semiclassical approximation with  backreaction. We show  that the end-point of the evaporation process  is AdS$_2$ with a vanishing dilaton, i.e. a regular, singularity-free, zero-entropy state. We explicitly compute the entanglement entropies of the black hole and the radiation, as functions of the horizon radius,  using   the conformal field theory (CFT) dual to AdS$_2$  gravity. 
We use  a simplified  toy model, in which  evaporation  is described  by the forming and growing of  a negative mass configuration in the   positive-mass black hole interior. This is similar to the ``islands'' proposal,  recently put forward to explain the Page curve for evaporating black holes.
The resulting  Page  curve for   AdS$_2$ black holes is in agreement  with unitary evolution.  The entanglement entropy of the radiation initially grows, closely following a thermal behavior, reaches a maximum at half-way of the evaporation process, and then goes down to zero, following the Bekenstein-Hawking entropy of the black hole. 
Consistency of our simplified model requires  a non-trivial identification of  the central charge of the CFT describing AdS$_2$  gravity with  the number of species of fields  describing  Hawking radiation. 
\end{abstract}
%


\newpage
\tableofcontents

%
%
%
%
\section{Introduction }
Since the discovery of  Hawking radiation  \cite{Hawking:1974sw, Hawking:1974rv}, the information paradox for an evaporating black hole has been one of the most intriguing puzzles of fundamental  theoretical physics.  At the semiclassical level, unitarity of quantum mechanics seems to be lost when a black hole is formed from a collapsing pure quantum state and then completely  evaporates, leaving behind only thermal Hawking radiation \cite{Hawking:1976ra}, described by a mixed quantum state (see, e.g. Refs.~\cite{Polchinski:2016hrw, Harlow:2014yka, Mathur:2009hf} for reviews).

Over the years, several possibilities of addressing the problem have been put forward. Either information may be lost forever  \cite{Unruh:1995gn, Unruh:2017uaw} -  as firstly advocated by Hawking \cite{Hawking:1976ra}, whose argument, however, conflicted with energy conservation \cite{Banks:1983by}-  or one could have remnants \cite{Chen:2014jwq}  at the end of the evaporation, fuzzy structures at the horizon (fuzzball) \cite{Mathur:2005zp} or, finally, information leaks out and is somehow encoded in the Hawking radiation.

In  more recent times,  the  AdS/CFT  correspondence \cite{Maldacena:1997re, Gubser:1998bc, Witten:1998qj, Aharony:1999ti} and  the discussions triggered by the firewall argument  \cite{Almheiri:2012rt, Almheiri:2013hfa} supported   a solution  of the information  puzzle, which preserves the unitarity of quantum mechanics. On one hand,   the  AdS/CFT correspondence implies that any  gravitational bulk process, such as black hole evaporation, can be described in terms of a conformal field theory on the boundary, for which the evolution of quantum states is unitary.  On the other  hand,   it allows for a microscopic explanation  of the Bekenstein-Hawking black hole entropy (see, e.g. Refs.~\cite{Aharony:1999ti, Witten:1998zw, Strominger:1997eq, Cadoni:1998sg}), the backbone on which the  "Central Dogma"  \footnote{For an external observer, black hole quantum dynamics can be described as the unitary time evolution of $N\sim S_{BH}$ quantum states, where   $S_{BH}$  is the Bekenstein-Hawking black hole entropy \cite{Almheiri:2020cfm}.} of black hole information is based.

From the bulk perspective, one possible way to preserve unitarity of quantum mechanics during the evaporation is to assume that information leaks out from the black hole,  encoded in Hawking radiation. Using quite  general arguments of information  theory, Don  Page has shown   this can only happen at late times, after the so-called Page time, when roughly half of the black hole has evaporated away. This process is characterized by a well defined pattern for the entanglement entropy (EE) of the radiation - the famous Page curve \cite{Page:1993wv, Page:2013dx}.

Very  recent  developments  give further support  to information conservation during the black hole evaporation process. They are  based on a semiclassical  approach to compute the entropy of Hawking radiation  which consists in extremizing  an entropy functional, expressed as the sum of the contribution of the entropy of bulk fields and that of several, disconnected regions,  inside and outside the black hole, called ``islands'' \cite{Almheiri:2020cfm, Penington:2019kki, Almheiri:2019qdq}. The formula is manifestly a generalization of the holographic Ryu-Takayanagi formula \cite{Ryu:2006b, Engelhardt:2014gca} and is based on the idea of the ``entanglement wedge reconstruction'' \cite{Penington:2019npb, Almheiri:2019psf, Almheiri:2019hni, Almheiri:2019yqk},  which allows to semiclassically compute the Page curve by correctly keeping track of the entanglement structure of both the black hole and the radiation subsystems. At early times, before the Page time, the entanglement wedge of the black hole includes all the interior, with a  neat separation between radiation and  black hole degrees of freedom. The  entropy of the radiation will, therefore,  increase as Hawking quanta starts leaking out. At late times, after the Page time,   a contribution given by the islands, forming just behind the horizon, starts dominating.  This determines the consequent decrease of the Page curve, signalizing the purification of the final state of the radiation.

The new generalized entropy formula, however, does not tell  where the information is encoded and how it manages to escape from the horizon. This is mainly due to the fact that the formula is built in terms  of the low-energy gravitational theory and makes no reference to the underlying  microscopic theory and to the  would-be unitary  dynamics   of the $N\sim S_{BH}$ quantum states  building up the black hole.  Another drawback of the generalized entropy formula is that    computations  are in general quite difficult to perform.     

In view of this state of the art, it is quite important  to  consider gravitational systems, for which we have at least an effective description of the underlying microscopic dynamics and in which the semiclassical black hole dynamics is simple enough to allow for an explicit analysis. 

The most natural candidates are the two-dimensional (2D) AdS  (AdS$_2$) black holes of Jackiw-Teitelboim (JT) dilaton gravity  \cite{Jackiw:1984je, Teitelboim:1983ux, Grumiller:2002nm}. The latter represents one of the most studied $2D$ gravity model, as it allows to describe, otherwise more difficult to capture, features of four-dimensional gravitational black holes  \cite{Grumiller:2002nm, Cadoni:1993rn} and  has several features that make it suitable for the before-mentioned purposes.
Owing to  the fact  that Hawking radiation has a  purely  topological origin in these  models \cite{Cadoni:1994uf},  the semiclassical  dynamics of black hole solutions can be solved in closed analytical form. For small values of the  2D Newton constant $1/\phi_0$,  they allow for an effective description in  terms of  a dual CFT with central charge  $c=12 \phi_0$.  This enables  a microscopic derivation of the BH entropy \cite{Cadoni:1998sg}, so that the  Central Dogma is based on solid ground. The black  hole  spectrum  of the JT theory  contains  an energetically preferred, regular, ground  state  (the AdS$_2$  space-time), a state with zero temperature and zero entropy \cite{Cadoni:1998sg, Cadoni:2017dma}, which is the perfect candidate for the end point of  the evaporation process. Last but not least,  the dual CFT description  of JT black holes allows for an explicit computation of the black hole EE \cite{Cadoni:2007vf}.                    

These nice features allow to  describe  analytically the behavior of the entropy of both the black hole and the radiation and consequently to reconstruct the Page curve. In \cite{Almheiri:2019hni}, for example, this is done by using the  holographic entropy formula  \cite{Ryu:2006b, Hubeny:2007xt, Giataganas:2019wkd} and the entanglement wedge reconstruction idea, for a JT black hole coupled with a CFT$_2$ matter sector with a higher-dimensional holographic dual. In particular, the latter allows to connect the interior of the black hole with the radiation subsystem, in the spirit of the ER=EPR conjecture \cite{VanRaamsdonk:2010pw, Maldacena:2013xja, VanRaamsdonk:2009ar}. The same technique is applied in \cite{Gautason:2020tmk} to derive the Page curve in the 2D CGHS model \cite{Callan:1992rs, Russo:1992ax}. Finally, in \cite{Verheijden:2021yrb}, the Page curve for JT black holes is derived by considering the 2D gravitational model as the dimensional reduction of the 3D AdS gravity.

In this paper we tackle  the information problem for the JT black hole using an alternative approach. We investigate  the semiclassical dynamics of JT black holes, coupled to conformal matter, including the backreaction of the geometry, and  derive the Page curve for their EE in a closed form.  We  will do it without  using either   higher-dimensional theories or  the generalized holographic entropy formula.
 We show that, working in  the semiclassical approximation with backreaction, the end-point of the evaporation is the AdS$_2$ space-time, endowed with a vanishing dilaton, i.e. a regular, singularity free, zero entropy space-time. This suggests unitary evolution  and information conservation. 

 We proceed  by computing  the entanglement entropies of the black hole and the radiation, as functions of the horizon radius. This is done using  the  effective description of  AdS$_2$ quantum gravity  in terms of the  dual CFT, i.e in the  large central charge regime,  $c\gg1$, of the CFT.
The computation of the EE of the radiation is performed using a simplified  toy model, in which  black hole evaporation  is described  by the formation and growth of a negative mass configuration in the   positive-mass black hole interior. This setup  represents  a rough, simplified version of  the earlier-mentioned ``islands'' conjecture, which has been recently proposed to explain the Page curve for evaporating black holes (see, e.g. Refs.~\cite{Almheiri:2020cfm, Penington:2019kki, Almheiri:2019qdq, Gautason:2020tmk, Verheijden:2021yrb, Goto:2020wnk, Marolf:2020rpm, Kim:2021gzd, Hollowood:2020cou, Anegawa:2020ezn, Bousso:2021sji}).  Unlike the semiclassical entropy formula used in the aforementioned papers, the EE formula used in this paper allows us to capture also contributions of \textit{purely} quantum mechanical correlations between the interior and the exterior of the black hole, in line with the ER=EPR spirit as well. Moreover, these correlations arise in a natural and simple way in our model, without resorting to higher-dimensional duals. 

In the final part of our paper, we  compare the curve of the EE for Hawking radiation with those pertaining to the thermal entropy of the radiation and the Bekenstein-Hawking entropy of the black hole. The resulting  Page  curve for   JT black holes is in agreement  with unitary evolution.  The entanglement entropy of the radiation initially grows, closely following a thermal behavior, reaches a maximum at the half-way point, and then goes down to zero, closely following the Bekenstein-Hawking entropy of the black hole during the final stages of the evaporation process. 
Basic principle of thermodynamics, together  with   the existence of a dual CFT  description, imply a non-trivial identification of  the central charge of the CFT describing AdS$_2$  gravity with  the number of species of fields  describing Hawking radiation.\\

The structure of this paper  is  as follows.
In  Sect.~\ref{2DAdSblackholes}  we briefly review the classical and semiclassical properties of  JT black holes coupled  with conformal matter,  focusing on the conformal anomaly and backreaction  effects of the geometry. 
We investigate the semiclassical dynamics of the model in Sect.~\ref{Sec:Unitarity of black hole evaporation}, by considering  the evaporation process, both  in static coordinates   and in terms of boundary dynamics.  
The calculation of the EE associated with  the JT black hole  of Ref.~\cite{Cadoni:2007vf} is reviewed  in Sect.~ \ref{entanglemententropyofbh}.
In  Sect. \ref{Sec:Pagecurve} we  discuss  the information flow during the black hole evaporation and present the main results of this paper concerning  the Page curve for the JT black hole. We  compare the EE of the radiation with the thermal entropies of the radiation  and of the black hole.  We also derive the relationship between the central charge of the CFT dual to AdS$_2$  gravity  and the number of species of matter fields  in the Hawking radiation.  
Finally, in  Sect.~\ref{conclusions}  we state our conclusions.

\section{2D AdS black holes}
\lb{2DAdSblackholes}
In  this paper we consider  2D AdS black holes.  The simplest gravity model allowing for  this kind of solutions is JT gravity, described by the action
\be\lb{JTnoanomaly}
\mathcal{S}_{\text{JT}}=\frac{1}{2\pi} \int d^2 x \sqrt{-g} \ \phi \left(R+2\Lambda^2\right) + \mathcal{S}_{\text{matter}},
\ee
where $\phi$ is a scalar field (the dilaton), playing the role of the inverse coordinate-dependent 2D Newton constant,  $R$ is the 2D Ricci scalar, $\Lambda^2$ the  cosmological constant and $\mathcal{S}_{\text{matter}}$ is the action for  matter fields.  A slight generalization of this model has been proposed by Almheiri and Polchinski (AP) \cite{Almheiri:2014cka}, by  adding a constant term $\alpha$ to the dilaton potential:
\be\lb{ap}
 V(\phi)=  2 \Lambda^2 ( \alpha -\phi  ).
 \ee
Thus, the JT model can be considered as a particular  case ($\alpha= 0$) of the AP model. We  will make use  of this  feature in the following Section.
\subsection{Classical solutions in absence of matter}
\label{Sec:ClassicalVacuumSolutions}

In  absence of matter, JT gravity admits asymptotically AdS$_2$   black holes as solutions  \cite{Cadoni:1994uf}. In the Schwarzschild gauge  the metric and the dilaton read
\be\lb{staticblackholesolSchw}
ds^2 =-\left(\Lambda^2r^2-a^2 \right)dt^2+\left(\Lambda^2 r^2-a^2 \right)^{-1}dr^2,\qquad \phi(r) = \phi_0 \Lambda r,
\ee
where $\phi_0$ and  $a^2$ are integration constants, related to the Arnowitt-Deser-Misner (ADM) mass of the solution:
\be\lb{ADMmass}
M= \frac{1}{2}\phi_0 a^2 \Lambda= \frac{1}{2}\phi_0\Lambda^3  r_h^2.
\ee
$a^2 >0$, $a^2=0$ and $a^2<0$ represent, respectively, a space-time with a positive ADM mass, with Killing  horizon at $r = r_h = a/\Lambda$, the AdS vacuum (zero ADM mass) and a space-time with a negative ADM mass. Adopting the nomenclature of Ref.~\cite{Cadoni:1994uf}, we will refer to these as AdS$_+$ ($a^2>0$), AdS$_0$ ($a^2=0$) and AdS$_-$ ($a^2<0$). These three solutions represent different parametrizations, covering different regions, of the same manifold, as they are connected with each other by coordinate transformations. This means that the \textit{local} properties of the space-time described by the metric~(\ref{staticblackholesolSchw}) are the same, independently from the value of $a^2$. Therefore, the three spacetimes can be maximally extended  to obtain full AdS$_2$, which has no horizon and is geodesically complete. Nevertheless, AdS$_+$ can be interpreted as a  2D black hole  with an event horizon at $r = r_h$, if one takes into account  the physical meaning  of the  dilaton as the (coordinate dependent)  inverse 2D Newton  constant, demanding $\phi\ge 0$. This requirement    implies the existence of a   space-time  singularity   at $r=0$, where the 2D Newton constant  diverges. The $\phi=0$ line  has to be considered  as an inner boundary of the space-time, whose existence    allows one to  consider AdS$_+$ as a black hole, AdS$_-$   as a space-time containing naked singularities and AdS$_0$ as the ground state,   zero mass solution   \cite{Cadoni:1994uf}
\footnote{A similar conclusion can be reached if one considers the JT model as originated  from  spherical dimensional reduction of  3D  Ba\~nados-Teitelboim-Zanelli (BTZ) \cite{Achucarro:1993fd}  or higher dimensional \cite{Cadoni:1994uf} black holes. In this case, the positivity condition  $\phi\ge 0$ is required by the identification of $\phi$ with the radius of the compactified sphere.  }. 

Once we interpret AdS$_+$ as a black hole space-time, it is natural to associate thermodynamic properties to it, i.e. a temperature $T_H$ and an entropy $S_{BH}$:
\begin{subequations}
\begin{align}
&T_H = \frac{a\Lambda}{2\pi} =\frac{r_h \Lambda^2}{2\pi}=\frac{1}{2\pi} \sqrt{\frac{2M\Lambda}{\phi_0}}; \lb{Temperature}\\
&S_{BH} =2\pi\Lambda\phi_0 r_h= 4\pi  \sqrt{\frac{\phi_0 M}{2\Lambda}} = 2\pi \phi \left(r_{h} \right). \lb{Entropy}
\end{align}
\end{subequations}

The  AdS$_0$  vacuum solution with a  linear  dilaton, given by Eq.~(\ref{staticblackholesolSchw}) with  $a=0$, termed linear dilaton vacuum (LDV) in Ref.~\cite{Cadoni:2017dma}, is not the only  $M=0$ vacuum of the model.
The  theory allows also for full AdS$_2$ space-time solution endowed with a constant, identically vanishing,  dilaton,
\be\lb{CDV}
\phi = 0,
\ee
which has been called the constant dilaton vacuum (CDV) solution in Ref.~\cite{Cadoni:2017dma} \footnote{Similarly, the AP model admits both AdS black hole solutions with a linear varying  and a constant non-vanishing dilaton \cite{Almheiri:2014cka}.}.
 
At first sight, the CDV and the LDV  seem degenerate in energy (they both have  zero ADM mass),  but  closer inspection  reveals that they are  separated  by a  mass gap $M_{\text{gap}}=\Lambda/(2\pi^2 \phi_0)$ \cite{Cadoni:2017dma, Almheiri:2014cka}.  Indeed the CDV, being full AdS$_2$ space-time,  does not admit finite energy excitation \cite{Maldacena:1998uz}.  The issue can be better understood if we  consider the  JT model  as a particular case     of the AP model, where   the two vacua are connected by an  interpolating solution  $\phi = \alpha^2 + \phi_0 \Lambda r$.  
Following Ref.~\cite{Cadoni:2017dma}, we can now  remove the apparent degeneracy  between the CDV and the LDV and study the thermodynamic behavior of the two vacua by looking at the free energy difference $\Delta F = F^{\text{LDV}}-F^{\text{CDV}}$. 
At $T_H \neq 0$, $\Delta F = -\frac{2\pi^2 \phi_0}{\Lambda}T_H^2 <0$ \cite{Cadoni:2017dma}, which tells us  that the LDV is thermodynamically favoured. However, at $T=0$, for the CDV ($\phi_0 \to  0$), $M_{\text{gap}}$ diverges and one has $\Delta F = \frac{\alpha^4 \Lambda}{2\phi_0} \to \infty$ \cite{Cadoni:2017dma}. The  LDV is therefore not thermodynamically stable at zero temperature: a phase transition occurs, which drives the system down to the CDV.\\

Let us now describe  the black hole solution in the conformal gauge, using light-cone coordinates $x^{\pm}$,
\be\lb{conformalgauge}
ds^2 = -e^{2\rho \left(x^+, x^- \right)}dx^+ dx^-, \qquad x^{\pm} = x^0 \pm x^1,
\ee
which will be  also  used  to describe the coupling of our model to matter fields. The field equations and constraints, stemming from the action~(\ref{JTnoanomaly}), in absence of matter, read now
\begin{subequations}
\begin{align}
&\partial_+ \partial_- \rho = -\frac{\Lambda^2}{4}e^{2\rho}; \lb{noback1}\\
&\partial_+ \partial_- \phi = -\frac{\Lambda^2}{2}\phi e^{2\rho};\lb{noback2}\\
&\partial^2_+\phi - 2\partial_+ \phi\partial_+ \rho =0; \lb{noback3}\\
&\partial^2_-\phi - 2\partial_- \phi\partial_- \rho =0. \lb{noback4}
\end{align}
\end{subequations}
They are solved by
\begin{subequations}
\begin{align}
&e^{2\rho} = \frac{4}{\Lambda^2}\left(x^--x^+ \right)^{-2};\lb{conformalfactorsolution} \\
&\phi= \frac{b+c\left(x^++x^- \right)+dx^+x^-}{x^--x^+ }, \lb{f4}
\end{align}
\label{vacsolutionslightcone}
\end{subequations}
where $a, b, c$ are constants.  The metric part of the solution~(\ref{conformalfactorsolution}) has an  SL(2,R) isometry, i.e it is invariant under the transformations:
\begin{equation}
x^{\pm}\rightarrow \frac{\alpha x^{\pm}+\beta}{\gamma x^{\pm}+\delta}, \quad \text{with} \ \alpha \delta-\beta \gamma=1.
\label{SL2R}
\end{equation}
By exploiting this invariance, the dilaton~(\ref{f4}) can be recast as  \cite{Cadoni:1994uf}:
\be\lb{lineardilatonstandard}
\phi = \frac{2\phi_0}{\Lambda}\frac{1-\frac{a^2 \Lambda^2}{4}x^+x^-}{x^--x^+}.
\ee
The coordinate transformations relating the solutions  written in the    Schwarzschild gauge~(\ref{staticblackholesolSchw})  to those written in the the conformal gauge~(\ref{vacsolutionslightcone}) are:
\begin{subequations}
\begin{align}
&x^+ = \frac{2}{a\Lambda}\tanh \left\{\frac{a\Lambda}{2}t-\frac{1}{2}\operatorname{arcsinh}\left[\left(\frac{\Lambda^2r^2}{a^2}-1 \right)^{-1/2} \right]  \right\},\\
&x^- = \frac{2}{a\Lambda}\tanh \left\{\frac{a\Lambda}{2}t+\frac{1}{2}\operatorname{arcsinh}\left[\left(\frac{\Lambda^2r^2}{a^2}-1 \right)^{-1/2} \right]  \right\}.
\end{align}
\lb{fromLCtoSCHW}
\end{subequations}
The AdS$_0$ solution is easily obtained by taking the zero mass limit, $a \rightarrow 0$, of Eq.~(\ref{lineardilatonstandard})
\be\lb{LDVCC}
\phi = \frac{2\phi_0}{\Lambda} \left(x^--x^+ \right)^{-1},
\ee
while the metric conformal factor~(\ref{conformalfactorsolution}) remains unchanged. 
This is the consequence of a general and peculiar feature  of the JT theory, which is easily seen from the Eq.~(\ref{noback1}): the classical dynamics of the conformal factor is independent from the dilaton. This implies that the dynamics of the model  is fully  encoded in the scalar field, which determines the global properties of the space-time through the  evolution  of the $\phi=0$ space-time symgularity. As  we will show in the next Section,  this feature remains   true also when we couple the gravity model to conformal matter both at the classical and semiclassical  level.
\\

\begin{figure}
\centering
\begin{tikzpicture}

\node (I)   at (-4,0)   {};
\node (A) at (-1,3) {$\bf{x^+}$};
\node (B) at (-7,3) {$\bf{x^-}$};

\path  
  (I) +(90:4)  coordinate[label=90:{$x^+=x^- =2/a\Lambda$}]  (Itop)
       +(-90:4) coordinate[label=-90:{$x^+ = x^- = -2/a\Lambda$}] (Ibot)
       +(180:4) coordinate[] (Ileft)
       ;
\draw[thick] (Ileft) -- 
          node[midway, above left, sloped]    {$x^- = 2/a\Lambda$}
          node[midway, below, sloped] {}
   (Itop) --
           node[midway, below, right] {$x^+ = x^-$}
          (Ibot) --
          node[midway, above, sloped] {$x^+ = -2/a\Lambda$}
          node[midway, below left]    {}    
         (Ileft) -- cycle;

\draw[->]
(I) edge (A) (I) edge (B);

\end{tikzpicture}
\caption{Penrose diagram of 2D AdS$_+$. The future and past  asymptotic  singularities are highlighted, corresponding to the two vertices on the line $x^+ = x^-$ of the diagram. The two tilted  lines correspond to the future $x_H^- = 2/a\Lambda$ and past $x_H^+ = -2/a\Lambda$ event horizons.}
\label{SketchStatic}
\end{figure}
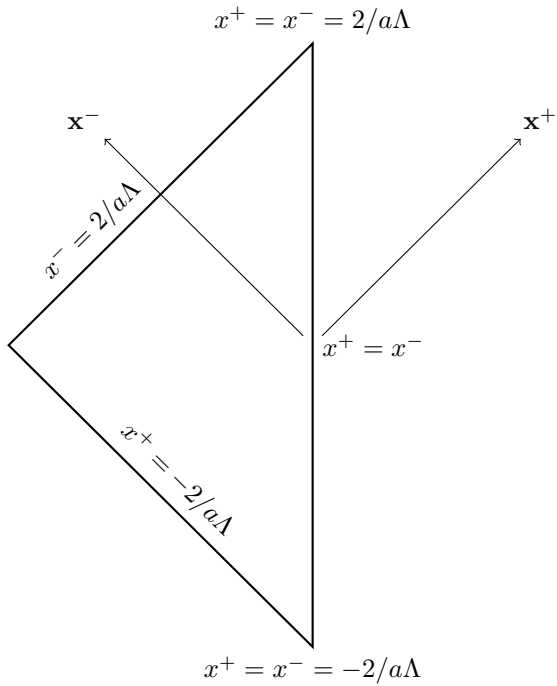

Let us briefly  discuss the  Penrose  diagram  of our space-time.
In light-cone coordinates, full AdS$_2$ space-time has two  disconnected parts, which are represented by two wedges in the  Penrose diagram. The   asymptotic conformal boundary $x^+=x^-$ of the space-time is  timelike. AdS$_+$ is half of  AdS$_2$, which in this paper is taken as the left wedge, following the convention  of Ref. \cite{Cadoni:1994uf} (see Fig.~\ref{SketchStatic}). This means that we are considering the region  $x^-\ge x^+$  of  AdS$_2$. The event horizon, $r = r_h = a/\Lambda$ in Schwarzschild coordinates, in light-cone coordinates corresponds to:
\begin{equation}
x_H^+ =-\frac{2}{a\Lambda}; \qquad x_H^- = \frac{2}{a\Lambda}.
\lb{EH}
\end{equation}
Considering the solution~(\ref{lineardilatonstandard}), the space-time singularity occurs at $1-\frac{a^2 \Lambda^2}{4}x^+x^-=0$.  The singularity is  always shielded by the event horizons~(\ref{EH}). The black hole interior corresponds to the region $x^-\ge 2/a\Lambda$ and  $x^+\le -2/a\Lambda$. If one uses light-cone coordinates,  so that  the scalar $\phi$ takes  the form~(\ref{f4}), the singularity is visible  for an asymptotic timelike observer  sitting on  the conformal  asymptotic space-time  boundary. He/she  
 will hit  the singularity at finite time,   $x^-= -2/a\Lambda$ (past singularity) and  $x^-= 2/a\Lambda$  (future singularity),  at least for finite non-vanishing value of $M$. However,  $\phi$ is not form-invariant under the isometric  SL(2, R) transformation (\ref{SL2R})  and the presence of the future  timelike  singularity in the asymptotic boundary can be avoided by choosing an appropriate light-cone frame, which removes  the $x^+x^-$ term in the dilaton solution~(\ref{f4}). 
In fact, using an appropriate SL(2, R) transformation,  the dilaton solution~(\ref{f4}) may be written as follows,
\begin{equation}
\phi = \frac{2\phi_0}{\Lambda}\frac{1+\frac{a\Lambda}{2}\left(x^-+x^+ \right)}{x^--x^+}.
\label{dilatonafterSL}
\end{equation}
In this new light-cone  frame, the singularity trajectory $\phi=0$ is 
\begin{equation}
x^- =-\frac{2}{ a\Lambda}-x^+,
\end{equation}
which gives only a past  singularity $x^- =-\frac{1}{a\Lambda}$, once evaluated on the timelike  asymptotic  boundary $x^- =x^+$. \ In Sect.~\ref{Sec:BoundaryDynamics} we will show that the SL(2, R) isometry of the metric  can also  be used to remove the asymptotic, future timelike singularity  in presence of  Hawking radiation and related  backreaction of the geometry.

\subsection{Coupling to matter, conformal anomaly and evaporation}
\lb{Sec:SemiclassicalBlackHole}
Let us now couple our  gravity model  to   matter  fields,    quantize the latter in the semiclassical approximation and include the backreaction of the geometry. \\
We consider  the matter sector in the form of $N$ massless scalar fields $f_i$, minimally coupled to gravity, described by the classical, conformally invariant,  action:
\be
\mathcal{S}_{\text{matter}} =-\frac{1}{4\pi}\int d^2x \sqrt{g} \  \sum_{i=1}^N \left(\nabla f_i \right)^2.
\ee
Quantization of the matter fields and backreaction of the geometry  is studied at the semiclassical level  by considering the quantization  of the CFT matter on the curved, classical,  2D gravitational background. This implies  a  non-zero trace of the (classically traceless)  stress-energy tensor for the matter fields (conformal anomaly) \cite{Christensen:1977jc}. For $N$ massless fields $f_i$ in two space-time dimensions, the conformal anomaly reads $\langle T^{\mu}_{\mu} \rangle = \frac{N}{12}R$, which can be accounted for by adding a non-local Polyakov-Liouville term in the JT action~(\ref{JTnoanomaly})
\be\lb{anomalyaction}
\mathcal{S}_{\text{anomaly}}=-\frac{N}{96\pi} \int d^2x \sqrt{g} \ R \Box^{-1} R,
\ee
where $\Box^{-1}$ is the inverse of the Laplacian. The field equations and the constraints, in light-cone coordinates  become local  and are given by 
\begin{subequations}
\lb{back}
\begin{align}
&\partial_+ \partial_- \rho = -\frac{\Lambda^2}{4}e^{2\rho}; \lb{back1}\\
&\partial_+ \partial_- \phi = -\frac{\Lambda^2}{2}\left(\phi -\frac{N}{24}\right)e^{2\rho};\lb{back2}\\
&\partial^2_+\phi - 2\partial_+ \phi\partial_+ \rho =-\frac{1}{2}\sum_{i=1}^N\partial_+ f_i\partial_+ f_i+\frac{N}{12}\left[\left(\partial_+ \rho \right)^2-\partial_+^2 \rho + t_+(x^+) \right]; \lb{back3}\\
&\partial^2_-\phi - 2\partial_- \phi\partial_- \rho =-\frac{1}{2}\sum_{i=1}^N\partial_- f_i\partial_- f_i+\frac{N}{12}\left[\left(\partial_- \rho \right)^2-\partial_-^2 \rho + t_-(x^-) \right]; \lb{back4}\\
&\partial_+ \partial_- f_i =0, \lb{back5}
\end{align}
\end{subequations}
where $t_{\pm}(x^\pm)$ are integration functions,  which have to be  determined by imposing appropriate boundary conditions.

As anticipated in the previous Section, we see that, also in the semiclassical treatment, the conformal factor of the metric $\rho$ is insensitive to the presence of matter, the dilaton  and backreaction effects. All dynamical information on the  evolution of the  semiclassical  system  is completely encoded in the solution  for the dilaton, which  determines the evolution of the space-time boundary at $\phi=0$. 

Another striking feature of Eqs.~(\ref{back}) is the fact that, at  the level of the field  equations, the conformal anomaly, i.e  semiclassical  quantum effects,  can be reabsorbed by  means of  a translation of the dilaton.  In fact, by performing the translation 
 \begin{equation}
\phi = \varphi + \frac{N}{24},
\label{dilatontransformation}
\end{equation}
and  using the solution of  Eq.~(\ref{back1}),  Eqs.~(\ref{back}) reduce to
\begin{subequations}
\lb{backt}
\begin{align}
&\partial_+ \partial_- \rho = -\frac{\Lambda^2}{4}e^{2\rho}; \lb{backt1}\\
&\partial_+ \partial_- \varphi = -\frac{\Lambda^2}{2}\varphi e^{2\rho};\lb{backt2}\\
&\partial^2_+\varphi - 2\partial_+ \varphi\partial_+ \rho =-\frac{1}{2}\sum_{i=1}^N\partial_+ f_i\partial_+ f_i+\frac{N}{12} t_+ ; \lb{backt3}\\
&\partial^2_-\varphi - 2\partial_- \varphi\partial_- \rho =-\frac{1}{2}\sum_{i=1}^N\partial_- f_i\partial_- f_i+\frac{N}{12} t_- ; \lb{backt4}\\
&\partial_+ \partial_- f_i =0 \lb{backt5},
\end{align}
\end{subequations}
which coincide with the classical ones~(\ref{noback2}), apart from the dilaton  translation and the presence  of the functions $t_{\pm}$.  The effect  of the conformal   anomaly is just  a translation of  the space-time boundary, which now  is located  at $\phi = N/24$ and the appearance  of the functions $t_{\pm}$  in the field equations.

These functions $t_{\pm}$  play an important role.  Their presence is a consequence  of the  anomalous  transformation law of $T_{\pm\pm}$,   which is given in terms of  the Schwarzian derivative of  light-cone transformation function  (see Eqs.~(\ref{fromLCtoSCHW})).
Usually  the $t_{\pm}$  are   fixed by imposing boundary   conditions  on Hawking radiation at past infinity (see, e.g. Ref.~\cite{Callan:1992rs}). 

Let  us conclude by noticing that the backreaction effects can  be reabsorbed by the translation~(\ref{dilatontransformation})  also at the level of the  action. 
This can be done by including  a purely  topological term $\Phi_0 R$ into the action~(\ref{JTnoanomaly}), with $\Phi_0$ constant,
and  performing the  shift $\Phi_0 = \hat\Phi_0 + \frac{N}{24}$ together with the translation~(\ref{dilatontransformation}).   

\section{Black hole evaporation}
\lb{Sec:Unitarity of black hole evaporation}

In the previous Section, we showed that  the classical and semiclassical dynamics of our model  is independent  from the metric, but is fully encoded in the solution for the dilaton.
We have therefore only two options to describe the evaporation process: $(1)$ we use a  static coordinate patch covering  the black hole exterior and we model  the evaporation as a succession of states of decreasing mass, or $(2)$  we use light-cone coordinates, in which the $\varphi=0$ singularity  is visible and  describe the evaporation process in terms  of boundary dynamics.\\
Using the results of Refs.~\cite{Cadoni:1994uf,Cadoni:2017dma} as a guide, we will find that in both cases  the end point of the  black hole evaporation process is  full, regular AdS$_2$  space-time endowed with a constant (vanishing) dilaton, i.e a singularity-free state with zero mass and zero entropy.
   
\subsection{Black hole evaporation  in the static patch} 
We  consider  for simplicity the evaporation of an initially static black hole of a given mass $M$. We do not consider the initial  phase of black hole formation from incoming matter, so that we can set  $f_i=0$. We model the  evaporation process  as  a sequence of static states of decreasing mass.   Following  Ref.~\cite{Cadoni:1994uf}, if one neglects  the backreaction,   black hole evaporation can be described as  a generalized Unruh effect \cite{Unruh:1976db}. Similarly to what happens in  quantizing    scalar fields in Rindler space-time, an AdS$_0$ observer  will detect the AdS$_+$ vacuum as filled with a  thermal flux of particles, with a Planckian spectrum at a temperature given by Eq.~(\ref{Temperature}).  The main difference between  Minkowski/Rindler and AdS$_0$/AdS$_+$      
space-times is that, in the latter case, thermal effects are not related to a physical  relative accelerated  motion of different observers, but have purely topological origin \cite{Cadoni:1994uf}. \\
\noindent The relevant equations are
\begin{subequations}
\begin{align}
&\rho'' = \frac{\Lambda^2}{4}e^{2\rho};\\
&\varphi'' = \frac{\Lambda^2}{2}e^{2\rho}\varphi;\\
&\varphi''-2\rho'\varphi' =\frac{N}{12}\left[\left(\rho' \right)^2-\rho'' \right],
\end{align}
\label{staticapproach}
\end{subequations}
where primes here stand for derivation with respect to the static coordinate $\sigma \equiv \frac{2}{a\Lambda}\operatorname{arcsinh}\left[\left(\frac{\Lambda^2 r^2}{a^2}-1 \right)^{-1/2} \right]$. \\
It is important to  notice that the new coordinate $-\infty<\sigma\le 0$  covers only  the  outside   horizon  region. In particular, the  space-time singularity at $\varphi=0$ is not visible. Thus, in this coordinate system, the backreaction of the geometry cannot be described by the boundary dynamics and is encoded instead in the change of the parameter $a$ (the black  hole mass).   

\noindent A set of solutions of Eqs.~(\ref{staticapproach}) is given by
\be\lb{AdSplusstatic}
e^{2\rho} = \frac{a^2}{\sinh^2\left(\frac{a\Lambda}{2}\sigma \right)}; \qquad \varphi = \frac{\phi_0 a}{2}\coth\left(\frac{a\Lambda}{2}\sigma \right)+\frac{N}{24}\left[ \frac{a\Lambda \sigma}{2} \coth\left(\frac{a\Lambda}{2}\sigma \right)-1\right],
\ee
describing the AdS$_+$ space-time, while 
\be\lb{lineardilatonvacuum}
e^{2\rho} = \frac{4}{\Lambda^2\sigma^2}; \qquad \varphi = \frac{\phi_0}{\Lambda\sigma},
\ee
corresponds to the AdS$_0$  LDV written in terms of the coordinate $\sigma$.

One can easily check that when the  black hole evaporates  and $M$  (hence the temperature~(\ref{Temperature})) decreases, the black hole interior region shrinks. When we take the  zero mass  limit, $a\rightarrow 0$, 
Eq.~(\ref{AdSplusstatic}) becomes  Eq.~(\ref{lineardilatonvacuum}). At first glance this seems to imply that  the AdS$_+$ black hole will settle down to the AdS$_0$ vacuum at the end of evaporation. However,  this is not actually  true, being  the  CDV energetically preferred,  according to the discussion of Sect.~\ref{Sec:ClassicalVacuumSolutions}. A phase transition will  bring the AdS$_0$ LDV to the AdS$_2$  CDV, with $\varphi=0$ (corresponding to $\phi=N/24$).  The end point  of the black hole evaporation  is therefore   full  AdS$_2$,  i.e a regular space-time with zero   mass and  entropy.  This  strongly suggests  that the evaporation of 2D AdS black holes is a unitary process.        \\

\subsection{Boundary dynamics}
\label{Sec:BoundaryDynamics}
In the previous Subsection   we have used a coordinate system covering only the black hole exterior,  in which  the inner space-time boundary, i.e the singularity, is not visible.  Here, we use light-cone coordinates, which also cover the black hole interior and allow us to describe  black hole evaporation in terms of  boundary dynamics. We  solve the field equations and the constraints, taking into account  the contributions of the Hawking flux  and the backreaction of the geometry. In the most  general case one should solve Eqs.~(\ref{backt}) by considering  a given profile of incoming matter, characterized by  $T_{++}$, and impose appropriate boundary conditions to fix the functions  $t_\pm$. Being the conformal factor of the  metric fixed, this should  allow to find the  solution for $\varphi(x^+,x^-)$, which gives the boundary equation when equated to zero. However, for our purposes we do not need to solve the  equations in such cumbersome detail. 

To keep the discussion as simple as possible we do not consider the initial phase of black hole formation from incoming matter,  but only the evaporation phase of an initially static black hole configuration  of given mass $M$.  Flux of Hawking radiation is switched on  at time $t=0$ in a non-adiabatic way. This allows us to set $\partial_+f_i=0$ and  $t_+=0$  in Eqs.~(\ref{backt})  for $t \geq 0$. Obviously, the form  of the stress-energy tensor component  $T_{--}$, which contains also information  on the incoming matter,  will be not determined in this way.   We will therefore  consider a general form for the stress-energy tensor $T_{--}$ of  Hawking radiation, which will be described by  a  generic function of $x^-$ only, $T_{--}=\tau_{--}(x^-)$. We will then show, using quite general   conditions on  $\tau_{--}(x^-)$ that the SL(2, R) isometry of the metric can always be used to remove   the future asymptotic   timelike singularity of the space-time. We will then provide an explicit description of the whole evaporation process by choosing a particularly simple form for  $\tau_{--}(x^-)$.

In order to solve the system~(\ref{backt}) and provide a simple analysis of the dynamics of the boundary, it is convenient to  introduce   a function $\mathcal{M}(x^-, x^+)$  parametrizing  the  field $\varphi$ \cite{Almheiri:2014cka},
\begin{equation}
\varphi = \frac{\mathcal{M}(x^-, x^+)}{x^--x^+}.
\label{Mdilaton}
\end{equation}

This allows us to rewrite Eqs.~(\ref{backt})  as follows
\begin{subequations}
\lb{backM}
\begin{align}
&e^{2\rho} = \frac{4}{\Lambda^2}\left(x^--x^+ \right)^{-2}; \lb{M1}\\
&\left(x^--x^+ \right)\partial_+\partial_- \mathcal{M} +\partial_- \mathcal{M} - \partial_+ \mathcal{M}=0;\lb{M2}\\
&\partial_+^2 \mathcal{M} =0 ; \lb{M3}\\
&\partial_-^2 \mathcal{M}(x^-, x^+) = -\left(x^--x^+ \right) \tau_{--} \lb{M4}
\end{align}
\end{subequations}
where $\tau_{--}$  is the stress-energy tensor.
As usual the dynamics  for the conformal factor  decouples, so  that we can solve the system~(\ref{backM}) for $\mathcal{M}$ with    $\rho$  given by Eq.~(\ref{M1}). We get
\begin{equation}
\mathcal{M}(x^-, x^+) = c_1 + c_2 \left(x^++x^- \right)-\left(x^--x^+ \right)\iint \tau_{--} dx^- + 2 \iiint \tau_{--} dx^-
\end{equation}
where  $c_{1,2}$  are integration constants. They can be fixed by requiring  the solution  for $\varphi$, given by Eq.~(\ref{Mdilaton}), to match the vacuum solution~(\ref{dilatonafterSL}), when the Hawking flux is turned off ($\tau_{--}=0$). This gives $c_1 = 2\phi_0/\Lambda, \, c_2 = \phi_0 a$  and the final form of the  solution  for $\varphi$  is
\begin{equation}
\label{generaltauvarphi}
\varphi = \frac{1}{(x^--x^+)}\left[\frac{2\phi_0}{\Lambda}  + \phi_0 a \left(x^++x^- \right)-\left(x^--x^+ \right)\iint \tau_{--} dx^- + 2 \iiint \tau_{--} dx^- \right].
\end{equation}

At the end of Sect.~\ref{Sec:ClassicalVacuumSolutions}, we have seen that the SL(2, R) isometry of the metric allows one to choose a light-cone frame in which the asymptotic observer does not see any singularity.  \\
Let us now show that this is still true even in the presence of Hawking radiation and backreaction effects. For the full solution~(\ref{generaltauvarphi}), the singularity trajectory $\varphi=0$ is described by:
\begin{equation}\lb{bound}
\frac{2\phi_0}{\Lambda} + \phi_0 a \left(x^++x^- \right)-\left(x^--x^+ \right)\iint \tau_{--} dx^- + 2 \iiint \tau_{--} dx^-=0.
\end{equation}
On the asymptotic  timelike boundary $x^-=x^+$  we have
\begin{equation}\label{boundarytrajgeneral}
 x^- = -\frac{1}{\phi_0 a}\left[\frac{\phi_0}{\Lambda} +\iiint \tau_{--}dx^- \right].
\end{equation}
When $\tau_{--}$ is a positive-definite function, as it should be in the case of Hawking radiation, the right hand side of Eq. (\ref{boundarytrajgeneral}) is always negative.  This implies, again, that we can always remove  the future singularity in the asymptotic timelike boundary using an appropriate  light-cone frame.\\

In order  to have an explicit description of the black hole evaporation  process, let us now  model it  in a simple way, as a sequence of steps in which the Hawking flux can be taken as constant. This allows us to fix $t_-=\text{constant}$ in  Eqs.~(\ref{backt4}), which is given in terms of a running  black hole mass $\hat M\le M$. $\hat M$ will decrease  during the evaporation process, with $\hat M\to 0$ as  the end point approaches. 
    
It is well known that the stress-energy tensor describing the outgoing Hawking radiation is given in terms of  the Schwarzian derivative of the static coordinate transformations~(\ref{fromLCtoSCHW})  connecting the Schwarzschild and the conformal gauges, and reads \cite{Cadoni:1994uf}:
\be \lb{HawTminus}
\langle 0 | T_{--} |0\rangle = \frac{N}{48}a^2 \Lambda^2.
\ee 
This equation determines the function  $\tau_{--}$ in Eq. (\ref{generaltauvarphi})  and allows to write explicitly   the singularity   equation (\ref{bound})
\begin{equation}\begin{split}\lb{solev}
&\frac{2\phi_0}{\Lambda \alpha^2}+ \sqrt{\frac{2\hat{M}\phi_0}{\Lambda}}\left(x^++x^- \right)-\left(x^--x^+ \right) \frac{N\hat{M}\Lambda}{24\phi_0}\left(\frac{\left(x^- \right)^2}{2}+\mathcal{C}_0x^-+\mathcal{C}_1 \right)+ \\
&\frac{N\hat{M}\Lambda}{12\phi_0}\left(\mathcal{C}_2+\mathcal{C}_1 x^- + \mathcal{C}_0 \frac{\left(x^- \right)^2}{2}+\frac{\left(x^- \right)^3}{6} \right)=0,
\end{split}\end{equation}
where we used the expression of the (running) ADM mass, i.e. $\hat{M} = \Lambda \phi_0 a^2/2$, while the $\mathcal{C}_i$s are integration constants. From Eq.~(\ref{HawTminus}), it immediately  follows
\begin{equation}\lb{boundCi}
\mathcal{C}_2 + \mathcal{C}_1 x^- + \mathcal{C}_0 \frac{\left(x^-\right)^2}{2}+\frac{\left(x^- \right)^3}{6} >0.
\end{equation}
Evaluating  Eq.~(\ref{solev})  on the asymptotic  boundary $x^+=x^-$ we get, 
\begin{equation}\begin{split}
f(x^-)\equiv\frac{2\phi_0}{\Lambda \alpha^2}+ 2\sqrt{\frac{2\hat{M}\phi_0}{\Lambda}}x^- +\frac{N\hat{M}\Lambda}{12\phi_0}\left(\mathcal{C}_2+\mathcal{C}_1 x^- + \mathcal{C}_0 \frac{\left(x^- \right)^2}{2}+\frac{\left(x^- \right)^3}{6}\right)=0.
\end{split}\end{equation}
As expected, the  function $f(x^-)$ is positive definite  so that there is no future asymptotic singularity.

It is important to stress  that $\hat M$ becomes  smaller and smaller as the evaporation proceeds. In our simplified picture, the apparent black hole horizon  shrinks  as the   evaporation proceeds and    the asymptotic  timelike observer will never  hit the  singularity.   At the  end of evaporation, we have $\hat M=0$   and  the Hawking  flux also vanishes. From Eq.~(\ref{solev}) follows that the space-time  boundary disappears, i.e the solution becomes the LDV  solution~(\ref{LDVCC}).   This also holds true for the general  solution~(\ref{generaltauvarphi}).   $\tau_{--}$  must be a monotonically decreasing function of $x^-$  and $a\to0$  at the end point of the evaporation process, implying that solution~(\ref{generaltauvarphi}) becomes the LDV given by Eq.~(\ref{LDVCC}).  

One  should keep in mind that, although the asymptotic observer does not encounter any future singularity during the evaporation process, the AdS$_0$ space-time is not geodesically complete,  as can be seen by the fact that radial null geodesics terminate at  finite length at $r\rightarrow 0$. This is of course due to the fact that we cut the space-time at $r=0$ (see Sect.~\ref{Sec:ClassicalVacuumSolutions}). Only its maximal extension, i.e. the full AdS$_2$ space-time, is singularity free and geodesically complete.
On the other hand, the stability argument of Ref.~\cite{Cadoni:2017dma}  can be used again to argue that the true end point of the evaporation process will be the  CDV  with $\phi_0=0$,  full AdS$_2$ space-time, consistently with unitarity. 

In the next  Sections, we will  confirm   this conclusion by  investigating the  evolution  of  entanglement  entropy  of Hawking radiation  during evaporation   and by  constructing its Page curve.

 \section{Entanglement entropy  of 2D AdS black holes}
 \lb{entanglemententropyofbh}
In the present and the following Sections, we tackle the information problem during the evaporation process  of a 2D JT black hole.  We will do this   by taking into account  the entanglement entropy of both  black hole and Hawking radiation.
The  peculiarities of  2D AdS gravity will allow us to  have a  precise, quantitative description of the EE  of the hole and of  Hawking radiation during the  entire evaporation process, which accounts for the information  flow between  them. \\

In two space-time dimensions, black hole entropy can be  fully  ascribed to quantum entanglement. This is  due to the fact that the 2D Newton constant  (parametrized by the dilaton)  is wholly induced by quantum fluctuations of the geometry. 
This can be shown by working in the AdS/CFT correspondence framework, which peculiarly has a non-holographic realization in  two space-time dimensions  in terms  of a dual  (chiral)  two-dimensional  CFT (CFT$_2$) living in the bulk \cite{Cadoni:2000fq, Cadoni:2000kr}. The  existence of this  dual quantum gravity theory  is a crucial ingredient  because it allows to compute the  EE of the  JT  black hole  in terms of the EE of the dual CFT$_2$ in the  curved 
gravitational background \cite{Cadoni:2007vf}
\begin{equation}
\lb{eq1}
S^{(bh)}_{ent}=\frac{c}{6} \ln \left(\frac{L}{\pi r_h}\sinh \frac{\pi r_h}{L}\right),
\end{equation}
where $c=12 \phi_0$ is the central charge of the CFT ($\phi_0$  plays the role of  the 2D inverse Newton constant) and  $L$, $r_h$ are the AdS length (related to the inverse of the cosmological constant, i.e. $L =\Lambda^{-1}$) and the black hole radius, respectively. Notice that Eq.~(\ref{eq1})  holds true  only when  we  are allowed to  use our effective description of AdS$_2$ quantum gravity in terms of the dual CFT$_2$  with  $c\gg1$, corresponding to the weakly-coupled regime of the gravitational theory, $\phi_0^{-1} \ll 1$.

The computations of Ref.~\cite{Cadoni:2007vf}, leading to Eq.~(\ref{eq1}),  are performed  using an Euclidean instanton and can therefore be  easily  extended  to the case of the AdS$_-$, i.e a black hole with "negative mass".  In the following, we will make use of this result  to  describe the EE  of the evaporating black hole interior in a simple way.

The  thermal,  Bekenstein-Hawking (BH)  entropy  $S_{BH}$ of AdS$_2$ black holes   can be derived as the leading term in the large mass expansion of the  EE~(\ref{eq1}).
In fact for large black holes, $r_h/L\gg 1$,  $S^{(bh)}_{ent}$ gives  the BH entropy with a subleading log term:
\be
\lb{eq2}
S^{(bh)}_{ent}\approx S_{BH}- 2\phi_0 \ln S_{BH}.
\ee
The  formula~(\ref{eq1}) describes the entanglement entropy of an eternal AdS black hole and is in agreement with several results, which appeared in the literature:
\begin{itemize}
\item
Classical space-time structure, in particular  its connectedness,   emerges out of quantum entanglement \cite{VanRaamsdonk:2010pw, Maldacena:2013xja, VanRaamsdonk:2009ar};  
\item
The BH entropy  has its origin in  the entanglement entropy  of the two edges of  maximally extended AdS$_2$ space-time when the degrees of freedom (DOF) in one edge are traced out \cite{Maldacena:2001kr}.  The result~(\ref{eq1}) is  a slightly  different realization of the idea proposed in Ref.~\cite{Maldacena:2001kr}, where the entanglement entropy is generated by two copies of a CFT in an initial entangled state and by using a thermo-field double. Eq.~(\ref{eq1}) is instead obtained using a single CFT defined in the maximally extended space-time by  tracing  the degrees of freedom (DOF) in half of it;
\item
Holographic entanglement entropy  formulas \cite{Ryu:2006b, Hubeny:2007xt, Giataganas:2019wkd} give the EE of maximally extended  AdS space-time  in terms of the area of co-dimension two minimal surfaces, which can  be  identified  with  the event horizon.
\end{itemize}
Altogether, these  results indicate that black holes are quantum  gravity objects (possibly at horizon scale) for which the relevant DOF are not localized   near the event horizon, as  a simple-minded  interpretation of the Bekenstein-Hawking   area law would suggest. The  horizon area dependence of the BH entropy, and hence its holographic nature,  should be therefore related  to the area scaling  law    of EE entropy in QFT.   

Equation~(\ref{eq1}), together  with the microscopic derivation of the  BH formula  given in  Ref.~\cite{Cadoni:1998sg}, gives a simple and intuitive characterization of the information  content of an eternal  JT black hole.  As shown in Ref.~\cite{Cadoni:1998sg}, the BH  entropy simply counts the microstates  of the CFT$_2$ which  are dual  to the black hole of mass $M$, whereas Eq.~(\ref{eq1})  tells us  that information is stored in the black hole in the  form of quantum correlations  localized in the black hole interior.   On the other hand, Eq.~(\ref{eq1})  does not reveal if and  how the information stored   
in the black hole comes out  during evaporation. In order  to  understand this aspect, we need to discuss the Page curve and the information    
flow for the evaporating JT black hole. This will be the subject of the next Section. 

\section{ Information flow  and the Page curve for evaporating JT black holes}
\lb{Sec:Pagecurve}
Let us now consider  the black hole evaporation process in terms of the information flow between the shrinking  black hole  interior and its exterior, and the information carried by  Hawking radiation. We will describe this in a quantitative way, by computing the EE entropy of the  black hole with its  exterior and the EE of Hawking radiation.  We will use  a simplified  description  of the  process in terms of a sequence of static  states characterized by  constant black hole radius  $r_h$, so that the black hole EE can be given as a  function of $r_h$ at any time. 

As seen in the previous Sections, this simplified picture is fully  justified  in the  JT gravity context, since the evaporation process, including backreaction   can be described  as  a sequence of  static states  (see Sect.~\ref{Sec:BoundaryDynamics}).
We will also consider the simplest case in which  the Hawking radiation is given by $N$  right-moving  species of 2D massless scalar fields $f_i$.
This  will result in a particularly symmetric situation in which  Hawking radiation is  treated  in the  same way as  AdS$_2$ quantum gravity, i.e a (chiral) CFT$_2$ with central charge $c=N$. Although this may not be the most general situation, it is simple  enough to tackle the conceptual puzzles  involved in the black hole evaporation.

\subsection{Hawking radiation} 
In order to  describe Hawking radiation, we  parametrize the 2D black hole geometry using, as usual,   two sets  of light-cone coordinates  defined in the previous Sections.  The coordinates $u=t+r,\,v=t-r$  pertain to the frame of  the  asymptotic observer, are expressed in terms of  its time $t$ and radial coordinate $r$  and cover the black hole exterior only.   The  coordinates $U=x^+,\,  V=x^-$  define, instead, the frame of the inertial observer falling through the black hole horizon, with a related  time coordinate $\tau= (x^+ +x^-)/2$.
Correspondingly, the asymptotic observer will expand    quantum fields   using   $b_\nu$, $b^\dagger_\nu$ modes of a given $t$-frequency  $\nu$,   
whereas the infalling observer  will    use $a_\mu$, $a^\dagger_\mu$   modes   of a given $\tau$-frequency  $\mu$.
The modes $b_\nu$ can be expressed  in terms of $a_\mu$, $a^\dagger_\mu$ by a Bogoliubov transformation. In this way   the $a$-vacuum $|a\rangle_a$ (the quantum  vacuum for the infalling  observer)  is seen as a bath with a thermal spectrum, at the black hole Hawking temperature $T_H$, by the asymptotic observer. 

The  modes $b_\nu$  are not enough  to calculate the EE entropy of  Hawking radiation, as they are only defined in  the outside region (region $I$). Since Hawking radiation is entangled with the black hole interior, we need to introduce the quantum field modes $\hat b_\nu$, which are defined  in the black hole interior (region $II$) \cite{Hawking:1974sw, Polchinski:2016hrw}. Denoting with $H$ the Hamiltonian of the full system,  the mode $b^\dagger_\nu$  raises the energy  by a quantum $\nu$: $[H, b^\dagger_\nu]=\nu b^\dagger_\nu$ (i.e. it creates a particle of energy $\nu$  in the outside region $I$), whereas $\hat b^\dagger_\nu$  lowers  the energy  by a quantum $\nu$: $[H, \hat b^\dagger_\nu]=-\nu \hat{b}^\dagger_\nu$ (i.e. it creates a particle of negative energy $-\nu$  in the interior  region $II$).  
One can easily show that the $a$-vacuum can be expressed in terms of $ b^\dagger_\nu$  and  $\hat b^\dagger_\nu$ as \cite{Polchinski:2016hrw}
\be
\lb{eq4}
|a\rangle_a=\mathcal{A} \exp \left(\int \frac{d\nu}{2\pi} e^{- \nu/T_H}  b^\dagger_\nu\hat b^\dagger_\nu \right) |0\rangle_{b, \hat{b}},
\ee
with $\cal{A}$ normalization factor and $|0\rangle_{b, \hat{b}}$ the $b$-vacuum.
This equation tells us that  the $b$-modes of Hawking radiation are entangled with the $\hat b$ modes in the black hole interior.
\subsection{The Page curve}
During the evaporation  process, the thermal entropy of a JT  black hole, of initial radius  $r_h=R_H$  and final  radius $r_h=0$,  decreases from the initial value $S_{BH}(R_H)=2\pi \phi_0 \Lambda R_H$ to  the final $S_{BH}(0)=0$.  Correspondingly, the thermal  entropy  of the Hawking radiation $S_R$, which  is roughly proportional to the number of quanta emitted, will grow  from   $S_R=0$ to $S_R\approx M_{BH}/T_H\approx S_{BH}(R_H)$.     
This is the essential of the information loss problem: assuming the JT black hole is formed by the collapse of a  quantum pure state, the evaporation process transforms a pure into a mixed quantum state.  

Conversely, if we assume that evolution is unitary  or, more precisely,  if we assume the validity of the so-called  "Central Dogma" \cite{Almheiri:2020cfm}, the quantum state of Hawking radiation has to be purified, and so its EE $S_E$  must go to zero in the final stages of the evaporation process.

Although there is no general consensus  about the mechanism  that purifies  the radiation, Page has shown, using  general 
principles of information theory, that information can only  come out  at late times. The result is the famous  Page curve for  $S_E$ \cite{Page:1993wv, Page:2013dx}. $S_E$  starts from zero  and initially grows, closely following the thermal entropy of the radiation  $S_R$, until the latter intersects $S_{BH}$  at approximately half-way of the evaporation process. At the intersection point $t=t_{Page}$, called Page time, $S_E$ reaches a maximum and then decreases, closely following  the  Bekenstein-Hawking entropy curve  $S_{BH}$  at late times. At the end of the evaporation, $S_E$ becomes zero again (the final state  is a pure state). 

\subsection{$2D$ Black hole Information}
A crucial issue in explaining  the Page curve is to understand the way information  can flow  from the black hole into late  Hawking radiation, so that the radiation final state  can be purified.  Recent  attempts, such as the wormhole and the island  proposal (see, e.g. Refs.~\cite{Almheiri:2020cfm, Penington:2019kki, Almheiri:2019qdq, Gautason:2020tmk, Verheijden:2021yrb, Goto:2020wnk, Marolf:2020rpm, Kim:2021gzd, Hollowood:2020cou, Anegawa:2020ezn, Bousso:2021sji}) have been focused  on  effects  of  the (Euclidean) low-energy  gravitational  theory, which may  be responsible for transferring information from the  evaporating black hole interior to the outside  radiation.
This  latter approach is not completely satisfactory because  it  leaves the question about the microscopic origin  of these low-energy  effects unanswered. Despite some  interesting proposals  (see, e.g. Refs.~\cite{Giddings:2011ks, Giddings:2012dh, Giddings:2012gc, Horowitz:2003he, Papadodimas:2012aq, Avery:2012tf, Verlinde:2013uja, Hooft:2016itl, Liu:2020gnp}), we are far away from having a  clear hint of how  the  $N\sim S_{BH}$ quantum states  building the black hole may evolve unitarily  during   black hole evaporation, transferring  completely the  information  of the initial  pure state, which collapsed to form the black hole,  into  late time  Hawking radiation.

The strategy we follow in this paper  is to use a simplified model   for  black hole evaporation, which allows to connect  the  microscopic  effective  description of  the black hole in terms of  a 2D  CFT with the evaporation process.  This is possible on account of  the  simplicity and  peculiarities  of 2D  JT gravity.
In fact, it has been observed that the  black  hole entropy can  be fully ascribed   to quantum entanglement if the Newton constant is induced by quantum fluctuations \cite{Fiola:1994ir}. The original version  of the proposal  referred to  quantum fluctuations of matter fields,   but,  in the context of the AdS/CFT correspondence in 2D, it has been extended to  the CFT  degrees of freedom dual to  2D AdS gravity.
The peculiarity of the AdS/CFT correspondence in 2D is the fact that it has a bulk/bulk realization,  in terms of a chiral CFT living in  2D  \cite{Cadoni:1998sg,Cadoni:2000fq, Cadoni:2000kr}.
It  follows that  thermodynamics and the evaporation  process of  the JT black hole allows for an effective description in terms of a 2D CFT with central  charge $c$ given by the inverse of 2D Newton constant  $\phi_0$  \cite{Cadoni:1998sg}: 
\beq\lb{e5a} 
c=12 \phi_0.
\eeq
In particular, this means that   black hole  entropy   has its origin in the quantum entanglement of $c$ microscopic DOF,  which give an  effective description of AdS$_2$ quantum gravity.   
$S^{(bh)}_{ent}$ equals the thermal entropy  $S_{BH}$ for large black holes, i.e when thermal fluctuations dominate. In this regime  we expect a semiclassical description to hold, i.e  the black hole to be described by a quantum, thermal, CFT of central charge $c$ in a classical gravitational background endowed with an event horizon.
Away from this semiclassical regime, we have contributions  coming from the quantum entanglement  of the microscopic  DOF. As expected, these corrections  are negative (see Eq.~(\ref{eq2})).  
In this generic quantum   gravity regime, we cannot simply describe the system as  a QFT in a fixed background geometry endowed with an event horizon. We expect   quantum  contributions  to the geometry to become relevant,  the classical  notion  of  horizon to loose much of its meaning and the inner structure of the black hole to play a role in the  black hole information problem.

These features are also evident in the computations of Ref.~\cite{Cadoni:2007vf} leading to Eq.~(\ref{eq1}). The EE is calculated  using an Euclidean instanton  and it  arises, similarly to Ref.~\cite{Maldacena:2001kr},   by tracing out  the CFT degrees of freedom over part  of the space. In this description, the BH entropy is not simply  the Boltzmann entropy of a CFT living in a boundary (the stretched horizon of  the black hole) of a structureless interior black hole space-time, but it is rather  due to quantum entanglement of  the black hole interior  with the outside world.
This change of perspective  also  implies that the holographic nature of the Bekenstein-Hawking formula has its roots  in  the  scaling of the EE with the area of the boundary separating  the observable from the unobservable region.

The interpretation of the black hole entropy as the semiclassical limit of   the EE entropy of some microscopic  quantum gravity (QG) DOF localized in the black hole interior  is also consistent  with the proposal of Ref.~\cite{Cadoni:2020mgb}, which sees  black hole thermodynamics as a manifestation of long-range quantum gravity effects.  In particular,   the (generalized) thermal equivalence  principle (GTEP) \cite{Tuveri:2019zor}, which is used to  explain the Hawking temperature, should be seen as a universal  property of  semiclassical horizons in the sense explained above.

Presently we do not have a precise formulation   of AdS$_2$ quantum gravity, but only a low-energy effective  description in terms of  a 2D CFT with central charge given by Eq.~(\ref{e5a}).  
We will  therefore use an extremely simplified  model to investigate the implications of our  quantum entanglement-based description  for the black hole information problem.  Our discussion  will remain quite general and based on general principles of QFT, so that we will not need to know an exact formulation  of  AdS$_2$ quantum gravity.  We will merely  assume that such a formulation does indeed exist.

\subsection{Entanglement entropy of Hawking radiation }
\label{Sec:EEHawkingRad}

In our simplified model, we describe black hole evaporation  as the emission of Hawking radiation (the modes $b_\nu$ of positive frequency $\nu$) in the exterior region  together with the  formation  of a state  of negative mass (the modes $\hat b_\nu$ of negative frequency $-\nu$) in the black hole interior.  
We have already seen that the two sets of modes  are entangled and can be considered as  two subsets,   $\mathcal{H}_b$, $\mathcal{H}_{\hat b}$ of the full Hilbert space spanned by  the modes $a$.
The  EE entropy of the radiation $S_E$ is the von Neumann entropy obtained by  partial tracing the full density matrix $\rho_{b\hat b}$ over $\mathcal{H}_{\hat b}$.  Well-known properties of  the EE imply that $S_E$ can be also calculated by partial tracing the density matrix over $\mathcal{H}_{ b}$
\beq\lb{e8}
S_E=- Tr \rho_{\hat b} \ln \rho_{\hat b},
\eeq
where $\rho_{\hat b}:= Tr_b \rho_{b\hat b}$.  

In order to simplify the calculation of  $S_E$, we will   assume that all the modes of negative frequency $-\nu$ will be localized in a connected  space-like slice   of size $\Sigma\le R_H$ of  the black hole interior, where $R_H$ is the initial black hole radius.  This is not necessarily true. We could also have  contributions   coming from   several disconnected  regions - islands using the terminology  of Refs.~\cite{Almheiri:2020cfm, Penington:2019kki, Almheiri:2019qdq, Goto:2020wnk, Marolf:2020rpm}.  However, we  do not expect this contribution to change  at least the qualitative behavior of our final results.  With this assumption we consider the localized,  connected   structure  generated  by  clumping the  $\hat b$-modes of negative energy as a 2D black hole of negative energy in  AdS$_2$, with $ \Sigma$ playing the role  of the  radius   of the negative mass black hole.
The entanglement entropy of an AdS$_2$ black hole with negative mass has been calculated in  Ref.~\cite{Cadoni:2007vf}. Using this result we get  
   
\beq
\lb{eq8}
S_{E}=\frac{N}{6} \ln \left(\frac{\mathcal{L}}{\pi\delta}\sin \frac{\pi \Sigma}{\mathcal{L}}\right),
\eeq
where $N$ is the number of species  of  fields and $\mathcal{L}$, $\delta$ are the IR, UV cutoffs respectively.
Being  $\Sigma\le R_H$,  we can identify the IR cutoff as the initial black hole radius: $\mathcal{L}=R_H$. Let us now denote  with $r_h$ the time-dependent value  of the black hole horizon during evaporation.  $r_h$ runs from $r_h=R_H$ at the beginning of evaporation to  $r_h=0$ at the end. Correspondingly, $\Sigma= R_H-r_h$ runs from $0$ to $R_H$. We get therefore  for the EE entropies of the black hole and  Hawking radiation

\beq
\lb{eq9}
S_{E}(r_h)=\frac{N}{6} \ln \left[\frac{R_H}{\pi \delta}\sin \frac{\pi (R_H-r_h)}{R_H}\right], \quad S^{(bh)}_{ent}(r_h)=\frac{c}{6} \ln \left(\frac{\delta}{\pi r_h}\sinh \frac{\pi r_h}{\delta}\right),
\eeq
where we have used the fact that  the UV cutoff is of order $L$ (see Ref.  \cite{Cadoni:2007vf}) and  set $\delta=L$ in Eq.~(\ref{eq1}).

The  entanglement  entropy~(\ref{eq9}) has well-known UV divergences caused by the contribution of  arbitrarily  short   wavelength modes.   In  our case these divergences manifest  themselves    at the beginning ($r_h=R_H$) and end point  ($r_h=0$) of evaporation, when $S_E$ blows up.   At these points  we have the leading logarithmic  divergences  
\begin{equation}\lb{ees}
S_E\simeq \frac{N}{6} \ln  \frac{r_h}{\delta}, \qquad S_E \simeq \frac{N}{6} \ln \left(\frac{R_H-r_h}{\delta} \right).
\end{equation}

\begin{figure}
\centering
  \includegraphics[width= 9 cm, height = 9 cm, keepaspectratio]{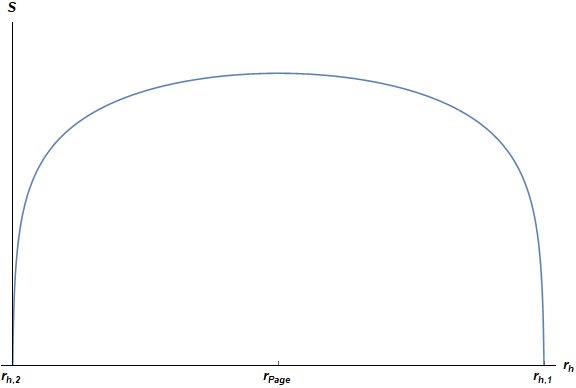}  
\caption{Regularized entanglement entropy $S_E^{(reg)}$ of the radiation  as a function of the black hole radius $r_h$, for the following  selected  values of the parameters: $N=1$ and $R_H/\delta =1000$. We plot  $S_E$ for $\delta\approx r_{h, 2}\le r_h \le r_{h, 1}\approx  R_H-\delta$. $S_E$ starts from zero at $r_h \approx \delta$, reaches its maximum at the Page radius  $r_h=r_{Page}= R_H/2$ when the black hole has reduced its size by  a factor $1/2$,  then decreases down to zero at the end of the evaporation process, at $r_h \approx R_H-\delta$. This behavior is consistent with a unitary evaporation process and has the form of the Page curve. Note that time runs towards decreasing values of $r_h$. \\}
\label{FIG:OnlyPagecurve}
\end{figure}

The UV  regulator  $\delta$ of the dual CFT  can be used to remove these divergences just by cutting off  $r_h$ at  distances above  $R_H-\delta$ and below $\delta$. The  regularized EE  $S_E^{(reg)}$  can then be obtained just by   
cutting the curve at the  points  $ r_{h, 2}=\frac{R_H}{\pi}\text{arcsin}\left(\frac{\pi \delta}{R_H} \right)$ and    $r_{h, 1} = \frac{R_H}{\pi}\left[ \pi- \text{arcsin}\left(\frac{\pi \delta}{R_H} \right)\right]$ where $S_E$ vanishes  (see Fig. \ref{FIG:OnlyPagecurve}). For small values of $\delta/R_H$, the intersection points become $r_{h, 2}\approx \delta$ and  $r_{h, 1}\approx  R_H-\delta$.

Alternatively, we can subtract the  finite $S_E(r_h=\delta)$  term from $S_E$ in such a way that the regularized EE  of radiation  $S_E^{(reg)}= S_E(r_h) - S_E(r_h=\delta)$  always vanishes exactly at $r_h=\delta$ and $r_h=R_H-\delta$.   
Notice that the  term  we  are   subtracting 
$S_E(r_h=\delta) = \frac{N}{6} \ln \left(\frac{R_H}{\pi \delta}\sin \frac{\pi \delta}{R_H} \right)$
consistently vanishes in the limit $\delta/R_H\to 0$.

\subsubsection{Thermal entropy of Hawking radiation and relation between $N$ and $c$}

The thermal entropy $S_R$ of Hawking radiation gives an upper  bound for its entanglement entropy  $S_E$ and characterizes  the thermodynamical regime of the radiation, where thermal  correlations  dominate over the quantum  ones. This allows for a coarse grained description of the radiation in terms of a thermal density matrix. The computation of $S_R$ and  the use of simple thermodynamical arguments will also allow  us to find a relation between  the central charge $c$ of the CFT    describing the JT black hole and the number of field species   $N$ in the Hawking  radiation.  
 
We derive the thermal entropy of the Hawking radiation at large temperature by first computing the von Neumann entropy of a single  mode  in a thermal bath at temperature $T$  and then integrating over the total number of modes.  We  put the system in a 1D box of  finite size  $\ell$, which therefore  acts as an IR cutoff.   The spectrum  for the    eigenvalues  $E_m$  of the Hamiltonian of the system  will be therefore discrete. At large $T$,  we can take $E_m= m \omega$, where  $m$ is a (positive) integer and  $\omega$  is of order $1/\ell$.  \\
The state~(\ref{eq4}) corresponds to a thermal density matrix $\rho$ of a single mode
\begin{equation}
\rho_{mm'} = \delta_{mm'} \frac{e^{-\beta E_{m}}}{\mathcal{Z}},
\label{densitymatrix}
\end{equation}
where $\beta = 1/T$, with $T$ temperature of the thermal bath in thermal equilibrium with  the hole.
$\mathcal{Z}$ is the partition function
\begin{equation}
\mathcal{Z} = \Tr\left(e^{-\beta \hat{H}}\right)=\sum_{m=0}^{\infty} e^{-m\omega/T} = \frac{e^{\omega/T}}{e^{\omega/T}-1}= \left(1-e^{-\omega/T} \right)^{-1},
\end{equation}
where $\hat H $ is the Hamiltonian of the system. Since we are considering $T\gg0$, we can neglect the contribution of the vacuum.

The normalized eigenvalues of the density matrix~(\ref{densitymatrix}) thus are:
\begin{equation}
p_m = \left(1-e^{-\omega/T} \right) e^{-m \omega/T}.
\end{equation}
It is easy to check that the normalization condition $\Tr \rho=1$ is satisfied.
The corresponding von Neumann entropy of the mode is:
\begin{equation}\begin{split}
S_{\omega} &= -\sum_{m=0}^{\infty} p_m \ln p_m =\frac{\omega/T}{e^{\omega/T}-1}-\ln\left(1-e^{-\omega/T} \right).\\
\end{split}\end{equation}

To compute the total entropy, we need to integrate over the number of modes $\mathcal{N}(\omega)$. If we consider the 1D volume $\ell$, each mode has wavenumber $k = m\pi/\ell$, where $m$ is again a positive integer. But $k=\omega$, so $m = \omega \ell/\pi$. The total number of modes in the $m$-space is  $\mathcal{N} = N \frac{1}{2} \cdot 2m = N m$,
where $2m$ is the volume of the  1D box, the factor $1/2$ accounts for the fact that $m$ is positive and $N$ for the number of species of fields.\\
In terms of $\omega$, $\mathcal{N}(\omega) = N \frac{\ell}{\pi}\omega$, and thus we have
\begin{equation}\begin{split}
S_R&=\int S_{\omega}\  d\mathcal{N}(\omega)=N\frac{\ell}{\pi}  \int_0^{\infty} \left[\frac{\omega/T}{e^{\omega/T}-1}-\ln\left(1-e^{-\omega/T} \right) \right]d\omega= N\frac{\pi}{3}T \ \ell.\\
\label{thermalentropyrad}
\end{split}\end{equation}

As  expected, the coarse-grained thermal entropy of the Hawking  radiation is given by the thermal entropy for a 2D CFT (massless bosons) on the plane with IR  regulator  $\ell$ and central charge  given by: 
\begin{equation}\lb{cchawk}
c_{HR}= N.
\end{equation}
As Hawking radiation lives in the AdS$_2$ background, the most natural choice for the IR regulator $\ell$ is to be of the order of the  AdS length $L$. In the following, we will  therefore set $\ell=\pi L$. 

At first sight, the use of $L$  both as the UV regulator  $\delta$ of the CFT dual to AdS$_2$ quantum gravity and as  the IR regulator   $\ell$  for Hawking radiation in the  AdS$_2$ background  seems contradictory. This apparent contradiction   can be solved by  the  UV/IR connection  in the context of the AdS/CFT correspondence \cite{Susskind:1998dq}, which relates the IR cutoff of the gravity theory with  the UV cutoff of the dual CFT. \\

So far the central charge~(\ref{e5a}) of the   CFT    describing  AdS$_2$ quantum gravity and the central charge~(\ref{cchawk}) of the CFT describing the Hawking radiation are two completely  independent quantities.  Let us now use  a  standard,  thermodynamical   argument to show that  $c=c_{HR}$.
Let's consider the  emission  of  an infinitesimal amount of energy  $dE_q$  from a black hole of radius $r_h$  and corresponding    mass $M$ and temperature $T_H$ given by Eqs. (\ref{ADMmass}) and (\ref{Temperature}) as a  reversible process.  After this emission, the black hole mass decreases to $M-dE_q$, its radius decreases to $r_h-dr_h$, whereas  the  black hole entropy decreases by the amount 
$dS_q \equiv \frac{dE_q}{T_H}$. Conservation of energy implies:
\begin{equation}
\frac{\phi_0}{2L^3}\left(r_h-dr_h \right)^2 +dE_q= \frac{\phi_0}{2L^3}r_h^2. 
\end{equation}
Neglecting terms of order $\mathcal{O}\left(dr_h^2\right)$ and using Eq.~(\ref{Temperature}) for the black hole   temperature we have \begin{equation}
dE_q = \frac{\phi_0}{L^3}\left(2\pi L^2 \right)^2 T_H \ dT_H .
\label{energyquanta}
\end{equation}
This gives the entropy change for the hole: 
 \begin{equation}
 dS_q =\frac{dE_q}{T}= 4\pi^2 \phi_0 L \ dT_H.
\end{equation}
Equating  this change of thermal entropy with  that of radiation, given by Eq. (\ref{thermalentropyrad}),  we get   
\begin{equation} \label{NC}
N = 12 \phi_0= c,
\end{equation}
where we used Eq. (\ref{e5a}).
This is  a  non-trivial relation between the number of species of fields $N$ in the Hawking radiation  and the central charge $c$ of the CFT dual to  the  AdS$_2$ quantum gravity.
The  result is  a consequence  of  the interplay between thermodynamics  and  field-theoretical features (in particular the AdS/CFT correspondence). Specifically, it follows from  $(1)$ the  reversible transfer of  coarse grained  entropy  from the black hole to the radiation  \cite{Alonso-Serrano:2015bcr, Muck:2016stv} $dS_{BH}=-dS_R$, $(2)$ the  description of both the black hole and the Hawking radiation in terms of a 2D CFT  and the related     linear scaling~(\ref{thermalentropyrad})  of the entropy with the temperature,    which allows to write  the number of species  $N$ in the Hawking radiation in terms of the (inverse of) 2D Newton constant $\phi_0$ and $(3)$ the AdS/CFT correspondence which implies Eq. (\ref{e5a}), i.e. it allows to write the central charge of the CFT dual to AdS$_2$ gravity in terms of $\phi_0$ .   

It is also interesting to notice that the derivation above is fully consistent with a corpuscular description of the black hole \cite{Cadoni:2020mgb}, i.e. its  description in terms of a bound state of $n$  quanta of  $N=c$  number of species,  with energy of the order of the temperature of the black hole, i.e. $E_q \sim T =r_h/2\pi L^2$. The corresponding infinitesimal change of the black hole mass  when $n$ quanta are emitted is therefore $dM \sim E_q dn \Rightarrow dn \sim dM/E_q \sim \frac{2\pi L^2}{r_h}\ dM$. But $dM \sim \frac{N}{12 L^3}r_h \ dr_h$, and hence:
\begin{equation}
dn \sim  \frac{\pi}{6 L}N dr_h \Rightarrow n \sim  \frac{\pi}{6 L}N r_h = S_{BH}.
\end{equation}
In the corpuscular description, Hawking evaporation  is just the transfer of energy and entropy from $n$ quanta building  the black hole to the $n$ Hawking radiation  quanta. Again, energy and entropy conservation require the number of species building  the black hole  to be the same of that composing  Hawking radiation.

\subsubsection{Entanglement entropy  and Page curve for the JT black hole}
\lb{pc}

In the previous Sections we have computed  all  quantities  characterizing the black hole and Hawking radiation, which are relevant from the  point of view of quantum information and thermodynamics, namely  $S_{BH}$ (Eq.~(\ref{Entropy})), $S_{R}$ (Eq. ~(\ref{thermalentropyrad})), $S_E^{(reg)}$ (left expression in Eq.~(\ref{eq9}))
\begin{subequations}
\begin{align}
&S_R = \frac{\pi c}{6} \frac{R_H-r_h}{\delta};\\
&S_E^{(reg)} = \frac{c}{6} \ln \left(\frac{R_H}{\pi \delta }\sin \frac{\pi r_h}{R_H} \right)\lb{sex} ;\\
&S_{BH} = \frac{\pi c}{6} \frac{r_h}{\delta}.
\end{align}
\end{subequations} 
We can now plot and discuss the  Page curve for the JT black hole.

\begin{figure}
\centering
\includegraphics[width= 11 cm, height = 11 cm, keepaspectratio]{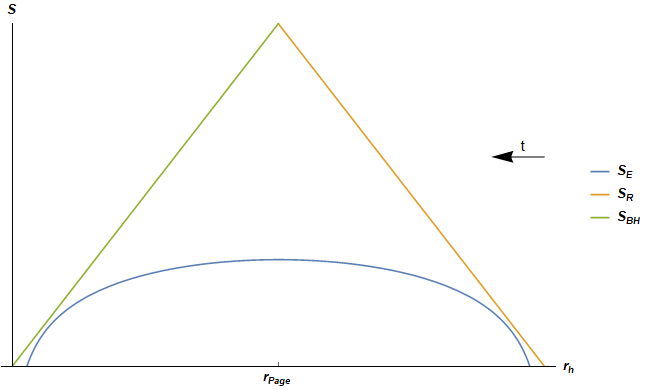}
\caption{Qualitative plot of the  entanglement entropy $S_E^{(reg)}$ of the radiation, thermal entropy of the radiation $S_R$ and the Bekenstein-Hawking entropy $S_{BH}$ as a function of $r_h$.  We show the curves for the following  selected  values of the parameters: $R_H=100$, $N=1$ and $\delta =20$. As $\delta$ approaches zero, the two zeros of $S_E^{(reg)}$ approaches to $r_h =0$ and $r_h = R_H$.  Note that the time runs towards decreasing values of $r_h$. }
\label{FIG:Entropies}
\end{figure}

The Page curve for the JT black hole is shown in Fig.~\ref{FIG:Entropies}, where it is  compared with the thermal entropy of Hawking radiation and to the Bekenstein-Hawking entropy of the black hole. 
Note that time increases towards decreasing values of $r_h$, from the beginning $t=0$ to the ending $t=t_E$ of evaporation. This corresponds to $r_h$ decreasing from the initial regularized black hole radius $r_h= R_H-\delta$ to the final regularized radius $r_h=\delta$. The plots show  a behavior of $S_E^{reg}$  which is fully consistent with  unitary evolution during the evaporation process  and  has  the form of the  Page  curve.
The EE of radiation starts from zero at $r_h = \delta$, reaches its maximum when the black hole has reduced its size by  a factor  $1/2$, $r_h= R_H/2$,  then decreases monotonically down to zero  at the end of the evaporation process, at $r_h = R_H-\delta$.

Conversely, the EE of the black hole, which for large ($r_h\gg L$) black hole is well approximate by the thermodynamical entropy   $S_{BH}$, starts from its maximum value at  $r_h= R_H$,  then monotonically decreases  to zero  at  $r_h=0$.

The Bekenstein-Hawking entropy $S_{BH}$ and the thermal entropy of radiation $S_R$ (green and orange solid lines respectively) behave linearly  as a consequence of the   CFT  description  and of  the space-time dimensionality. 
The  regularized entanglement entropy  $S_E^{(reg)}$   closely tracks the  behavior of the  thermal entropy of the radiation $S_R$ and the Bekenstein-Hawking $S_{BH}$ curves   at the
beginning and at the end of evaporation respectively, while in the other stages is determined  by the  behavior of the $\ln\sin$ function.
This is because, initially, most of the correlations in the radiation have thermal nature. As the evaporation proceeds, $S_E^{(reg)}$ begins to deviate strongly from $S_R$. At half evaporation, at the Page time,  $r_h=R_H/2$ in our 2D case, $S_E$  reaches its maximum and  the correlations in the radiation  begin to extract information from the black hole.  At the end of the evaporation process, when most of the information has been extracted from the hole, $S_E$ catches up with $S_{BH}$ and becomes thermal again.\\
The  curve for $S_E^{(reg)}$  is  symmetric with respect to the  "Page  radius" $R_H/2$.  This seems, again, to be a consequence of the low, $D=2$, space-time dimensionality.   

It is important to stress again  that, thanks to the equality between $N$ and the central charge $c$  of the CFT dual to  AdS$_2$  gravity, Eq.~(\ref{NC}), the $S_E$ curve is always bounded from above by the coarse-grained entropies $S_R$ and $S_{BH}$. As in Page's argument \cite{Page:1993wv}, this is due to the fact that the latter two retain less information, being related to thermal states, representing thus the upper bound of the entanglement entropy.   
\section{Conclusions}
\lb{conclusions}
In this paper  we have investigated  the semiclassical dynamics of  2D dilatonic JT black holes and derived  the Page curve for their entanglement entropy.  Our results are fully consistent with unitarity of the evaporation process.  Specifically, we  have shown that the end point of the evaporation is 2D AdS space-time  with vanishing  dilaton -  a  perfectly regular   state with zero mass and entropy.   We  have also  shown that, during the evaporation, the  behavior of the entanglement entropy of the radiation  agrees with Page's argument and with information preservation. In fact  the EE of the radiation  initially grows, following a thermal behavior, reaches a maximum at half-way of the process, and then goes down to zero, following the Bekenstein-Hawking entropy of the black hole. 
Moreover, the existence of a dual  CFT description for the JT black hole and usual thermodynamic arguments  imply a non-trivial identification of  the central charge of the CFT dual to  AdS$_2$  gravity with  the number of species of fields  in  the Hawking radiation.
 
 We have used a simplified model to discuss the semiclassical dynamics and to describe the black hole interior. In particular,  we  have  modelled the evaporation process by considering the contribution  of a single,  connected configuration of negative mass - or island, forming and growing inside the positive-mass black hole. Of course, contributions of several disconnected parts may be present.  Nonetheless, we expect  our simplified description  to give the leading contribution.

One nice feature of our  model  is that one of the main  assumptions  of the Central Dogma - the existence of a description  of the black hole in terms of $N \sim  S_{BH}$ quantum states -  certainly  applies.  Although  the precise mechanism that allows information to  escape from the black hole interior is not fully clear (this would require  a precise description of    AdS$_2$  quantum gravity), the  existence of an underlying  dual  CFT dynamics drastically improves our understanding of  the process of the information flow. On one hand, our outcome represents an independent confirmation of  several interesting results recently appeared in the literature \cite{Almheiri:2019hni, Gautason:2020tmk, Verheijden:2021yrb}, in which the Page curve for 2D black holes  is derived either using  higher-dimensional theories or an holographic  entropy formula. On the other hand, our model allows us to compute the Page curve for the EE entropy in a closed  form and in a rather simple manner, and to keep track of the quantum mechanical correlations between the interior and the exterior of the black hole in a natural way, by working entirely in the 2D theory context.

Another interesting feature of our model and our results is that they explain in a simple and intuitive way how geometry and quantum entanglement   are both essential to save unitarity in the black hole evaporation process. This happens  in a way which is fully consistent with the  ER=EPR proposal and the emergence of  the classical space-time structure out of quantum entanglement \cite{VanRaamsdonk:2009ar}. The informational   content of  the eternal JT black holes can be  considered  as the information contained in  one of the two edges of the maximally extended AdS$_2$ space-time \cite{Maldacena:2001kr}. As  evaporation   proceeds, after the Page time,  information extraction from the hole  occurs  in two steps.  During the first  one, which terminates with the black hole setting down to the AdS$_0$ vacuum,  only  quantum correlations in one  edge of AdS$_2$  are reconstructed.  In the second one, characterized by the phase transition from  AdS$_0$ to the CDV  (the two-edged  AdS$_2$),  the full quantum correlations in  the two edges  are restored, leaving behind a final pure state  for the radiation and  a regular space-time geometry. Unfortunately, our effective theory  cannot be used  to describe this second step, in particular to explain how quantum  correlations may emerge between classically disconnected regions of space-time. It is quite obvious  that, for these purposes, a full AdS$_2$ quantum gravity theory is needed. Actually, the phase transition by itself is most likely a signal  of the breakdown of our effective description and of the reorganization of  the  relevant  DOF.          

Apart  from its simplicity, our model has also other drawbacks. It is a 2D  model  with AdS asymptotics and it is not clear to what extent it may capture features of   asymptotically  flat  4D black holes.  The JT model appears in a variety of cases  as a description  of the near-extremal near-horizon regime of  charged and rotating 3D, 4D (and also D-dimensional) black holes (see, e.g. Refs.~\cite{Cadoni:1993rn, Cadoni:1994uf, Achucarro:1993fd, Giddings:1992kn, Trivedi:1992vh, Almheiri:2016fws, Nayak:2018qej, Moitra:2018jqs, Moitra:2019bub}).  Thus, our  model is a good approximation  for this kind of  black holes in the aforementioned  regime, but  its validity for generic   black holes has to be  further  investigated.

\bibliography{2DBlackHolesPageARXIV} 
\bibliographystyle{ieeetr}

\end{document}